# All-carbon vertical van der Waals heterostructures:
# Non-destructive functionalization of graphene for electronic applications

Mirosław Woszczyna[1], Andreas Winter[2], Miriam Grothe[1], Annika Willunat[2], Stefan Wundrack[1], Rainer Stosch[1], Thomas Weimann[1], Franz Ahlers[1], and Andrey Turchanin[1,2*]

[1]*Physikalisch-Technische Bundesanstalt, 38116 Braunschweig, Germany*
[2]*Faculty of Physics, University of Bielefeld, 33615 Bielefeld, Germany*

e-mail: turchanin@physik.uni-bielefeld.de
Tel.: +49-521-1065376
Fax: +49-521-1066002

## Abstract

We present a route to non-destructive functionalization of graphene via assembly of vertical all-carbon van der Waals heterostructures. To this end, we employ single-layer graphene (SLG) sheets grown by low-pressure methane CVD on Cu foils and large-area dielectric ~1 nm thick amino-terminated carbon nanomembranes ($NH_2$-CNMs) generated by electron-beam-induced crosslinking of aromatic self-assembled monolayers. We encapsulate SLG sheets on oxidized silicon wafers with $NH_2$-CNMs via mechanical stacking and characterize structural, chemical and electronic properties of the formed heterostructures by Raman spectroscopy and X-ray photoelectron spectroscopy as well as by electric and electromagnetic transport measurements. We show that functional amino groups are brought in close vicinity of the SLG sheets and that their transport characteristics are not impaired by this functionalization; moreover, we demonstrate a functional response of the heterostructure devices to the protonation of the amino groups in water. Due to its relative simplicity, the suggested approach opens broad avenues for implementations in graphene-based electronic devices where non-destructive chemical functionalization of graphene is required (e.g., for engineering electrical transducers for chemical and bio-sensing) or as complementary dielectric to graphene in hieratical heterostructures.

**TOC figure**

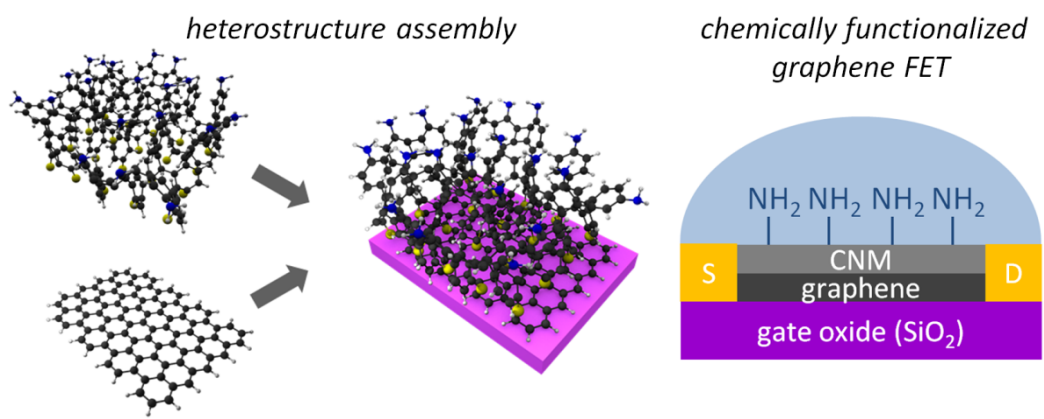

The functionalization of pristine graphene sheets is of key importance for their applications as electrical transducers in electronic devices such as, e.g., electric field-effect based nanosensors.[1,2] However, optimal routes to the functionalization have not been yet established which strongly restricts the applications.[3,4] As pristine graphene consists of exclusively $sp^2$-carbons organized in the honeycomb network, it is chemically relatively inert and difficult to functionalize via covalent bonding. Even when the covalent functionalization is achieved, due to associated structural modifications in graphene, its attractive electronic properties (e.g., high electrical charge mobility, strong ambipolar electric field effect, high thermal conductivity) are largely diminished.[1-3,5] Both covalent bonding to the graphene defects, as in oxidized graphene[6] or to graphene grain boundaries,[7] and direct bonding to the intact benzene rings[8] were studied and have demonstrated these limitations. In this respect, non-covalent functionalization of graphene, i.e. via weak van der Waals (vdW) forces, may provide an attractive alternative, as it does not induce severe changes into graphene.[4] Thus, the functionalization of graphene with flat polyaromatic molecules like porphyrins[9] has been shown. Although the adsorbed monolayers do not disrupt the graphene structure, their stability is low and limits implementations in devices. Therefore, a more durable functionalization of graphene based on the vdW interactions is highly desirable. Here we present a route to such functionalization via engineering of all-carbon vertical heterostructures by mechanical stacking[10] of *amino-terminated carbon nanomembrane* (NH$_2$-CNM) and *single-layer graphene* (SLG) sheets. CNMs are a novel two-dimensional (2D) carbon-based electronic material with dielectric properties made via electron-/photon-induced crosslinking of polycyclic aromatic self-assembled monolayers.[11,12] We employ SLG sheets, grown by low-pressure chemical vapor deposition (CVD) of methane on Cu foils[13], and amino-terminated ~1 nm thick CNMs[14-16], to assemble NH$_2$-CNM/SLG heterostructures on oxidized silicon wa-fers and use them for the fabrication of field-effect devices. The chemically active amino groups of $NH_2$-CNMs are located in these heterostructures in close vicinity to the graphene plane. We characterize structural, chemical and electronic properties of the large-area heterostructures and the fabricated electric-field devices by Raman spectroscopy, X-ray photoelectron spectroscopy (XPS) as well as by electric and electromagnetic transport measurements at room and low temperatures; and we study the electric field response of the devices in water. We show that the intrinsic electronic quality of pristine SLG sheets is preserved in the heterostructures, opening broad avenues for use in graphene-based electronic devices, e.g., for engineering electrical transducers for chemical and bio-sensing or as the complementary ultrathin dielectric material to SLG sheets for their integration with other materials.

A scheme for engineering the $NH_2$-CNM/SLG heterostructures on oxidized silicon wafers is presented in Figure 1. The process consists of two parallel routes including the fabrication of free-standing (i) $NH_2$-CNM and (ii) SLG sheets, which is finalized by their mechanical stacking into the vertical vdW heterostructure. To this end, $NH_2$-CNM sheets are prepared from self-assembled monolayers (SAMs) of 4'-nitro-1,1'-biphenyl-4-thiol on Au/mica substrates by electron irradiation (100 eV, 60 mC/cm$^2$) resulting in the lateral crosslinking of the biphenyl molecules[17] and the conversion of the terminal nitro groups into amino groups.[14,16] SLG sheets are grown by low-pressure CVD of methane on copper foils.[13] Then, the grown $NH_2$-CNM and SLG sheets are stacked on an oxidized highly doped silicon wafer by using the poly(methyl methacrylate) (PMMA) assisted transfer[18,19] (details in Supporting Information (SI)). In the formed heterostructure (see Fig. 1, right), amino groups of the $NH_2$-CNMs are in close vicinity to the graphene plane, as they are separated from its surface by an only ~1 nm thick dielectric sheet of cross-linked biphenylthiols.[18] They can further be flexibly chemically functionalized for applications.[16,20]

Fig. 2a shows an optical microscope image of the large-area $NH_2$-CNM/SLG heterostructure fabricated on an oxidized (oxide thickness ~300 nm) silicon wafer. Four different regions can clearly be recognized in this image corresponding to the bare substrate and the areas with $NH_2$-CNM, SLG, and $NH_2$-CNM/SLG heterostructure. As expected for the dielectric $NH_2$-CNM,[10] its optical contrast appears lower in comparison to the well-conducting SLG.[21] We employed Raman spectroscopy at ambient conditions to characterize graphene in different areas of this sample. As seen from Figure 2b, the intensity ratios between the characteristic D- and G- peaks of bare graphene and graphene integrated into the heterostructure are similar, which demonstrates that no additional structural defects are introduced into the SLG sheet upon stacking a $NH_2$-CNM on its top. Due to the disordered nature of $NH_2$-CNM and its monolayer thickness (~1nm), $NH_2$-CNM does not contribute of the measurable Raman intensity at our experimental conditions.[10] A detailed analysis of 15000 Raman spectra for both the bare SLG and $NH_2$-CNM/SLG heterostructure areas gave the same values for positions of G- and 2D- peaks, and for characteristic I(D)/I(G) and I(2D)/I(G) intensity ratios, which are 1589 $cm^{-1}$, 2680 $cm^{-1}$, 0.12 and 3.0, respectively (see supplementary information (SI), SI Table 1 and SI Figure 1 for details). We have found that the standard deviation of the full width at half maximum (FWHM) value of the G-peak for SLG in the heterostructure (FWHM=14.7±1.3 $cm^{-1}$) is a factor of two lower than for the bare SLG (FWHM=14.4±2.5 $cm^{-1}$). This difference is indicative for a lower degree of charged impurities[22,23] in the $NH_2$-CNM/SLG heterostructure in comparison to the bare SLG on silicon oxide. The concentrations obtained by electric transport measurements (see next paragraph) correlate very well with the spectroscopy data.

Figure 2c presents high resolution XPS spectra of the core level S2p, C1s and N1s electrons of the heterostructure and the bare graphene regions. Due to the presence of $NH_2$-CNM, the S2p and N1s signals are clearly detected for the heterostructure region, in which also an increase by 0.6 eV of the C1s FWHM value in comparison to the bare graphene is detected. The N1s signal is composed of two components at 399.3 eV and 401.2 eV, which are characteristic for pristine (-$NH_2$) and protonated (-$NH_3^+$) amino groups,[14] respectively. As determined from attenuation of the substrate Si2p signal (not shown) for the heterostructure and for the bare graphene, a $NH_2$-CNM contributes as expected with ~1.1 nm to the heterostructure thickness. A more detailed analysis of the XPS data is presented in SI Table 2. Thus, based on the spectroscopy characterization (Raman spectroscopy and XPS) we demonstrate that via the fabrication of the $NH_2$-CNM/SLG heterostructures terminal amino groups are brought in close vicinity of the SLG sheet (see Figure 1) without destroying its structural quality.

Next, we study by four-point measurements the transport characteristics of the SLG sheets integrated into heterostructures and compare them with the properties of bare graphene. To this end, a batch of large-area (140 μm × 25 μm) Hall bar devices with several side contacts (see Figure 3a) was fabricated from the same CVD grown SLG sheet placed on an oxidized silicon wafer. Then the wafer was sliced and a half of bare devices was directly examined, whereas on top of the second half a large-area $NH_2$-CNM was transferred to obtain heterostructure devices. In these devices, the $NH_2$-CNM covers both graphene areas and gold wiring with the bonding pads. Figure 3a presents an optical microscope image of one of the heterostructure devices. Ra-man mapping at positions of the characteristic D, G and 2D peaks (see Figure 3b) reveals its high structural quality and homogeneity on the large scale, which is directly confirmed by electric transport measurements. Figure 3c presents room temperature (RT) electric-field effect measurements as a function of back-gate voltage, $U_{BG}$, at four different side contacts of the device (see Figure 3a, P1-P4). As can be seen, the electrical characteristics are homogeneous on the scale of ~3500 $\mu m^2$. By keeping this device in high vacuum (~$10^{-5}$ mbar) at RT we observed a shift of the graphene charge neutrality point (CNP) from initially 10 V to 3.5 V after pumping for 18 hours. Since possible contaminations, trapped between the SLG sheet and the silicon oxide substrate and/or between SLG and $NH_2$-CNM, cannot be removed by this treatment, we attribute this change to desorption of the environmental adsorbates or the remaining lithography-resist rests from the outer surface of the device.

Figure 3d presents a comparison of the mobility data for heterostructure (H) and bare graphene (G) devices as a function of charge carrier concentration. Prior to the measurements, all samples were kept at least for two hours in high vacuum (~$10^{-5}$ mbar, RT). A striking difference is observed in the mobility data of H- and G-type devices. In the whole studied carrier concentration range (up to $n=5\times10^{12}$ cm$^{-2}$) H-devices have a higher mobility than G-devices. Thus for a hole concentration of $2\times10^{12}$ cm$^{-2}$ (see Figure 3e) the mobility of G-devices is $\mu$ = 1000–1600 cm$^2$/Vs, whereas H-devices have a mobility increased by about 80%, $\mu$ = 2500–2640 cm$^2$/Vs. We selected two devices of each type with the highest mobility – H3 and G1 – and studied their transport characteristics in more detail. Figure 3f shows the RT conductivity plots, $\sigma = \rho^{-1}$, as a function of carrier concentration in the vicinity of respective charge neutrality points ($U_{CNP}$ = 3.25 V for H3 and $U_{CNP}$ = 8 V for G1). By measuring a deviation of $\sigma(n)$ from the linear behavior, we estimate a residual carrier concen-tration, $n_*$, for both devices, which gives a measure for the variation of local electrochemical potential due to the charged impurities at graphene interfaces, the so-called electron-hole puddles, which may result from trapped charges in the silicon oxide or adsorbates at the silicon oxide/graphene or graphene/environment interfaces.[23] In this carrier concentration range both holes and electrons participate in transport. The extracted $n_*$ values are about 83×10$^{10}$ cm$^{-2}$ and 48×10$^{10}$ cm$^{-2}$ for G1 and H3 devices, respectively. These results are further confirmed by low-temperature measurements of the Hall coefficient, $R_H = \rho_{xy}/B\rho_{xx}$, in a magnetic field, $B$, of 1 T, where $\rho_{xy}$ and $\rho_{xx}$ are the Hall and the longitudinal resistivity. The single particle model predicts $R_H = 1/ne$ with a singularity at n = 0, whereas experimentally a smooth transition between electrons and holes is observed, representing a co-existence of both carrier types in the transition region.[24] Thus, in devices G1 and H3 we obtain mixing of the carrier concentrations in the range of 89×10$^{10}$ cm$^{-2}$ and 41×10$^{10}$ cm$^{-2}$, respectively. As at low carrier concentrations graphene mobility is limited by scattering on the charged impurities,[25] higher mobility values for H-devices in comparison to G-devices correlate with the lower residual carrier concentrations. It is peculiar that H-devices show a continuous decrease of the mobility with increasing carrier concentration beyond the electron-hole puddle ranges (see Figure 3d), whereas the mobility of G-devices stays nearly constant. We attribute this behavior to the encapsulation of graphene in H-devices with an about 1 nm thin dielectric sheet terminated with polar amino groups and sulfur species[16], which effectively screen[26] the charge impurities and suppresses scattering of the electrical carriers. A similar behavior was recently theoretically discussed for top-gated graphene devices with ultrathin gate dielectric layers.[27]

In the following we demonstrate that at low temperatures the magneto-transport properties of H-devices reproduce very well the quantum mechanical phenomena attributed to SLG, Figure 4. Thus, Figure 4a shows an evaluation of the Shubnikov-de Haas oscillations with their onset at about 3 T. From the oscillations period we find the hole concentration of $2.3 \times 10^{12}$ cm$^{-2}$, which perfectly corresponds to the value set by the back gate voltage $U_{BG}$. Moreover, at low magnetic fields a conductivity change due to the weak localization effect[28] is clearly observed (see inset to Figure 3a). We applied the weak localization theory[28,29] to perform fitting of these data (blue solid line) and to estimate scattering parameters. The obtained values for phase breaking, intervalley scattering and intravalley scattering times are $\tau_\varphi$ = 5.6 ps, $\tau_i$ = 0.5 ps, and $\tau_*$ = 6 fs, respectively; and the corresponding lengths are $l_\varphi$ = 370 nm, $l_i$ = 115 nm, and $l_*$ = 37 nm. These results are in good agreement with the scattering parameters of other transport studies of CVD graphene on silicon oxide substrates.[30] At a high magnetic field the quantum Hall effect (QHE) in device H3 appears as a sequence of plateaus accompanied by the longitudinal resistance oscillations (Figure 4b). The Hall conductivity (Figure 4c) follows a staircase shape with $4e^2/h$ high steps, as expected for SLG. Figure 4d presents the QHE in the vicinity of the filling factor $v$ = -2 in more detail. We observe a vanishing of the longitudinal resistance, which is indicated in a fully developed quantum Hall resistance plateau. In the range of minimum longitudinal resistance the mean value of the relative Hall resistance deviation, $\delta_r$, (see also figure caption to Figure 4d for details) yields a value of only $(-3.5\pm3.7)\times10^{-5}$. To the best of our knowledge, this is the most precise QHE measurement on CVD graphene reported so far. Thus, the magneto-transport measurements unambiguously demonstrate that the intrinsic quality of graphene is preserved in the fabricated NH$_2$-

CNM/SLG vdW heterostructure and that the chemical functionalization of SLG has been achieved in a non-destructive manner.

To test the heterostructures for possible sensor applications, we exposed H-devices to Millipore water at ambient conditions and measured their electrical response. $NH_2$-CNM sheets insulate the underlying graphene from water[11] and therefore the graphene resistivity can only be affected by a change in the electrostatic environment at the $NH_2$-CNM/water interface. A special care was taken to prevent the forming of water bridges between gold bonding wires of the devices (see SI for details). Water droplets (see inset to Figure 5b) were placed on the heterostructure area using a micropipette, and after the measurements they were then blown away by purging nitrogen. Figure 5a shows the ambipolar electric field effect of the same H-device at ambient conditions and in water. The presence of water results in *n*-doping of graphene and a shift of the CNP by about 27 V. We attribute this change to the protonation of amino groups (-$NH_3^+$) in $NH_2$-CNM, which result in an effective gating of graphene through the 1 nm thick dielectric layer of cross-linked biphenylthiols. As estimated from these data, a charge carrier injection into graphene corresponds to a carrier density change of ~$2×10^{12}$ $cm^{-2}$. With the dielectric constant of $NH_2$-CNMs ($\varepsilon \approx 2.9$, Ref.[10]) and using the plate capacitor model, we obtain an effective interfacial potential at the water/$NH_2$-CNM interface of ~120 mV (see SI, p. 5). Figure 5b shows the corresponding dynamic response of the device resistivity acquired by four point measurements with a direct current of 1 µA and $U_{BG}$=0 V. Just after placing the first water droplet (green arrow), which only partially covered the heterostructure area, a noticeable and rapid change of the resistivity was detected. After the full coverage with water (red arrow) we measured a giant increase by at least 400 % of the graphene resistivity, which corresponds to a decrease of the charge carrier concentra-tion in agreement with the ambipolar electric-field effect presented in Figure 5a. Blowing the droplet away (black arrow) results in a fast recovering of the resistivity to its initial value of ~1.6 kΩ/sq. Some instability of the resistivity signal in water was caused by the evaporation and droplet movement, which was observed in an optical microscope.

In summary we have demonstrated the fabrication and characterization of all-carbon $NH_2$-CNM/SLG heterostructures in which the terminal amino groups are separated by an only 1 nm thick carbon-based dielectric film from the plane of graphene. The intrinsic quality of the SLG sheets is not disturbed by this functionalization; moreover, $NH_2$-CNMs acts as the effective encapsulation layer improving the electric transport. As the amino groups can further be flexibly chemically functionalized[16,20], the suggested route opens broad avenues for the engineering of functional electronic devices (e.g., nanosensors). The electric response of the heterostructure devices to water shows the viability of this approach. As CNMs can economically be produced on large scale from a variety of aromatic monolayers,[12] it is promising to use them in the graphene-based electronics as complementary 2D dielectric, similar to hexagonal boron nitride sheets, for the engineering the efficient top-gate electrodes or for making the field-effect tunneling transistors.[31] An additional interesting opportunity is provided by the fact that CNMs can be converted into high electronic quality graphene by annealing.[32] In this way both graphene and insulating CNMs can be prepared from the same molecular precursor facilitating implementation of all-carbon electronics in nanotechnology.

**Supporting Information**

This material contains the detailed description of materials and methods, analysis of Raman spectroscopy and microscopy data, atomic force microscopy (AFM) images of bare graphene and heterostructure Hall bar devices, model calculations.

**Acknowledgements**

This work was supported by the Deutsche Forschungsgemeinschaft (SPP „Graphene", TU149/2-1, TU149/2-2 and WE3654/3-1; and Heisenberg Programme, TU149/3-1) and the EMRP project „GraphOhm" (The EMRP is jointly funded by the EMRP participating countries within EURAMET and European Union).

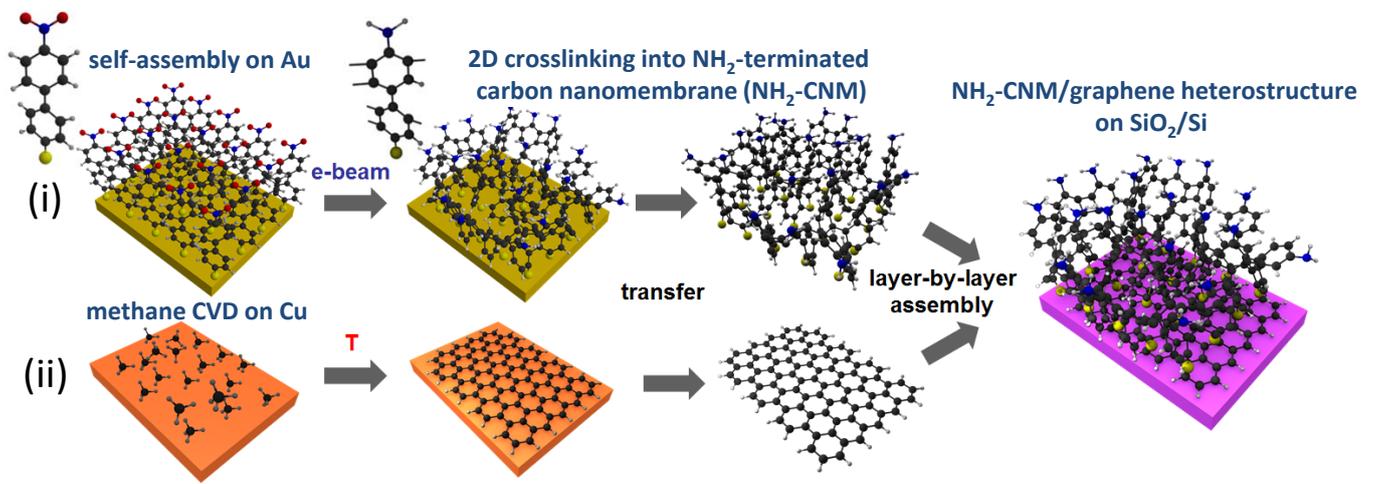

Figure 1

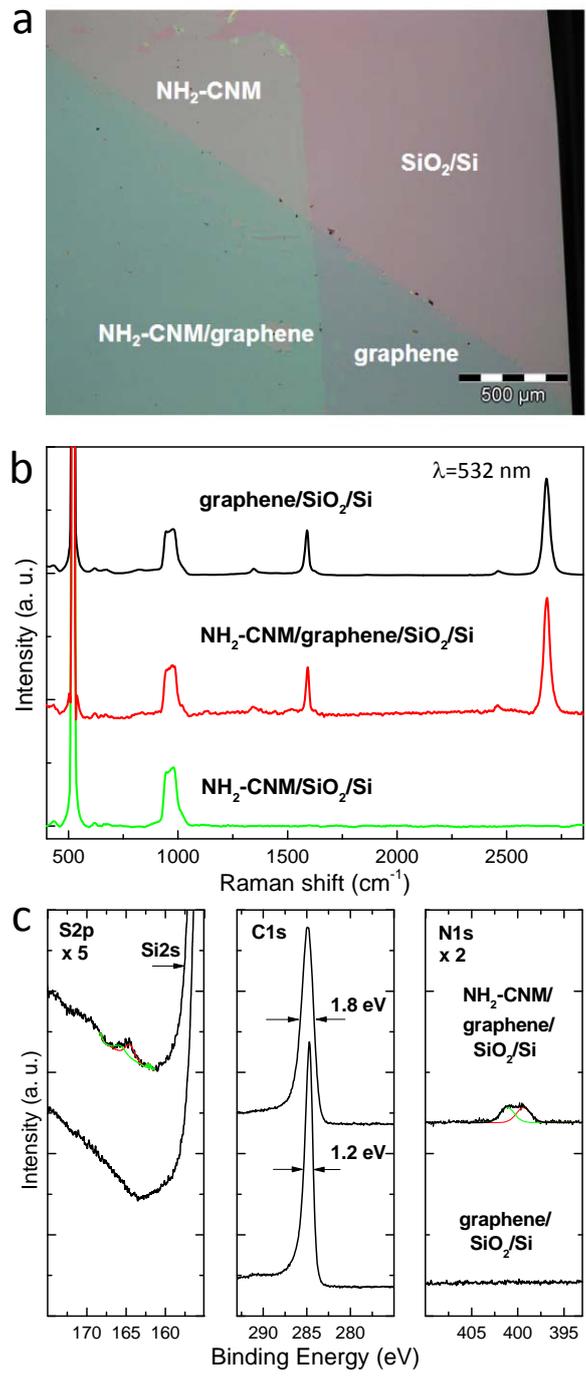

Figure 2

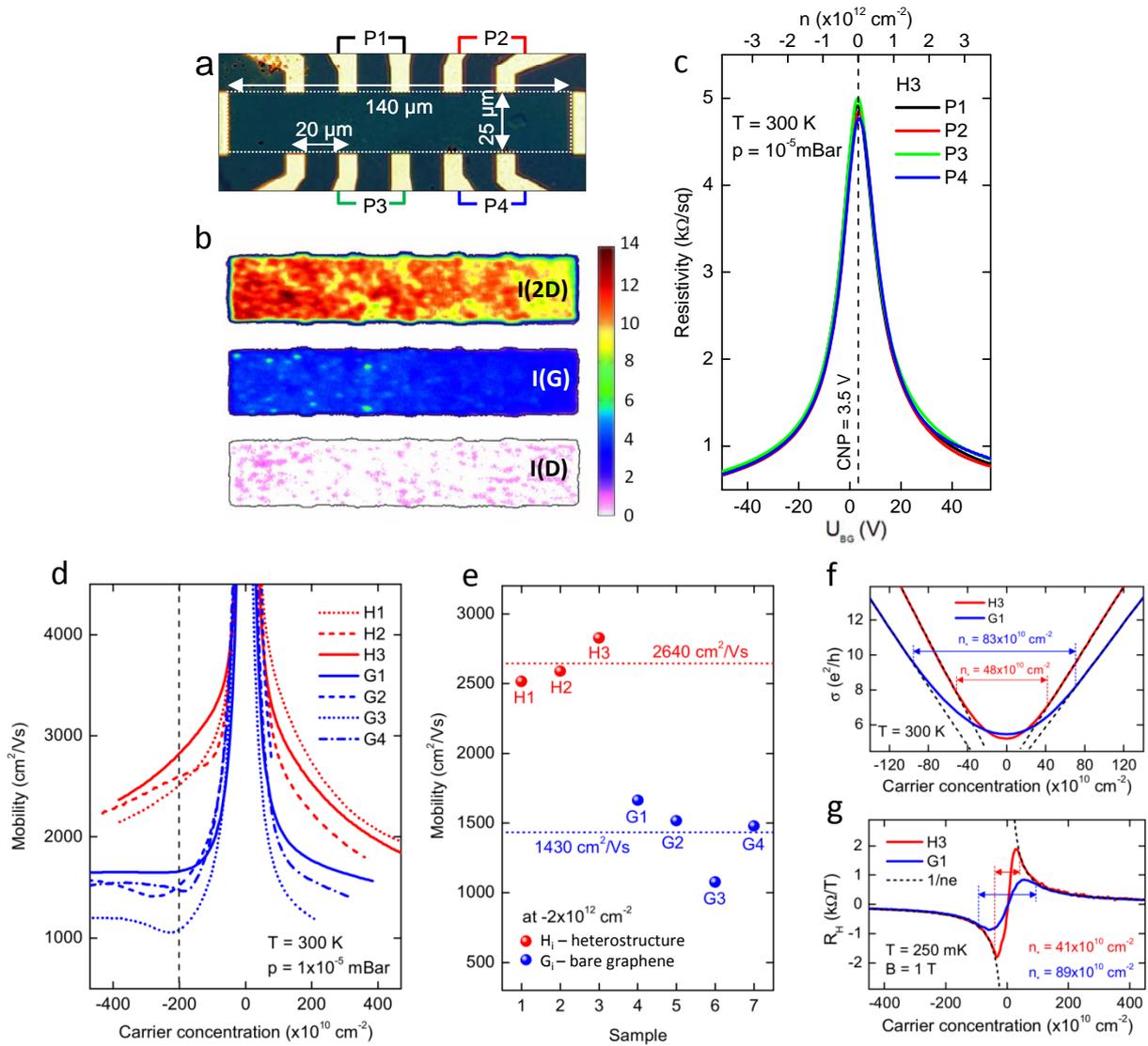

Figure 3

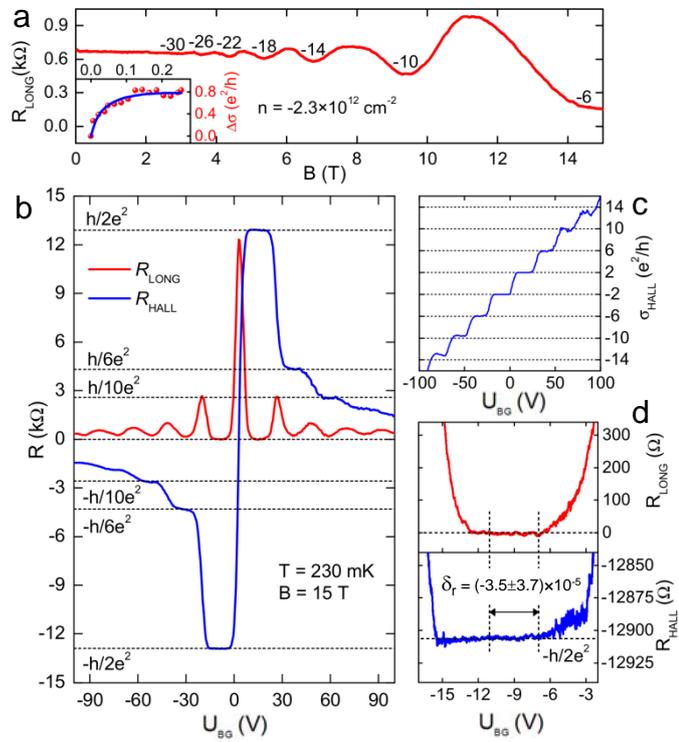

Figure 4

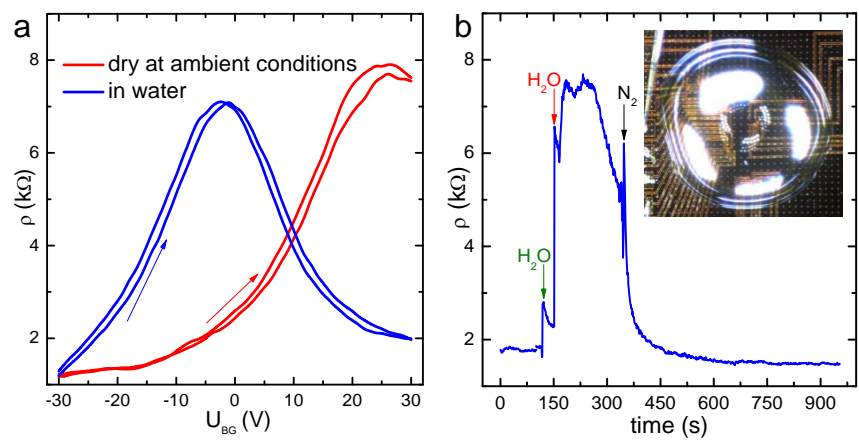

Figure 5

**Figure captions**

**Figure 1.** Scheme of the heterostructure assembly. In (i) amino-terminated carbon nanomembranes ($NH_2$-CNMs) are prepared from 4'-nitro-1,1'-biphenyl-4-thiol SAMs on Au/mica substrates by the electron irradiation induced crosslinking. In (ii) single-layer graphene (SLG) sheets are grown by CVD of methane on Cu foils. Finally, both types of sheets are separated from their original substrates and mechanically stacked to form the $NH_2$-CNMs/SLG heterostructure on an oxidized silicon wafer. Color code for atoms: black – carbon, grey – hydrogen, blue – nitrogen, yellow – sulfur, red – oxygen.

**Figure 2.** Characterization of structure and chemical composition of the heterostructure. **a,** Optical microscope image of the $NH_2$-CNMs/SLG heterostructure on an oxidized silicon wafer (300 nm oxide). Four different regions are recognized: bare $SiO_2$/Si; SLG; $NH_2$-CNMs; and $NH_2$-CNMs/SLG. The optical contrast was enhanced for better recognition. **b,** Raman spectra ($\lambda$ = 532 nm) characteristic of the SLG, $NH_2$-CNMs and $NH_2$-CNMs/SLG areas. **c**, X-ray photoelectron spectroscopy (XPS) data of the characteristic S2p, C1s and N1s peaks for the bare SLG and the $NH_2$-CNMs/SLG areas. The terminal amino groups (see N1s) are detected on the surface of the heterostructure.

**Figure 3.** Electric transport measurements of bare graphene (G) and heterostructure (H) electric field devices. **a**, Optical microscope image of the fabricated H-device. **b,** Raman spectroscopy maps ($\lambda$ = 532 nm) of the intensity of the characteristic 2D, G and D peaks. **c,** RT graphene resistivity at different side contacts (P1-P4) as a function of back-gate voltage, $U_{BG}$. **d**, RT electric charge mobility of graphene in H- and G-devices as a function of carrier concentration, $n$. **e**, Comparison of the RT electric charge mobility in H- and G-devices at $n$ = -2×$10^{12}$ cm$^{-2}$. **f**, RT conductivity of gra-

phene in H3 and G1 in the vicinity of their charge neutrality points (CNPs). **g**, Electrons and holes coexistence revealed by Hall coefficient measurements for devices H3 and G1 at a magnetic field of 1T and a temperature of 250 mK. Dashed lines correspond to the single-particle model for holes and electrons.

**Figure 4.** Magneto-transport measurements of the heterostructure device at low temperature (H3, T=230 mK). **a**, Shubnikov-de Haas oscillations in graphene with the corresponding filling factors as a function of magnetic field. The inset presents the weak localization effect in graphene at a low magnetic field. Data points are a conductivity change with respect to the zero magnetic field, whereas a blue solid line is a fitting curve from the weak-localization theory. **b**, The quantum Hall effect as a function of back-gate voltage at $B$ = 15 T. Red and blue plots show the longitudinal and Hall resistances, respectively. Horizontal dashed lines correspond to the theoretical quantum resistance plateaus for SLG. **c**, Hall conductivity calculated as $\sigma_{HALL} = \rho_{HALL}/(\rho_{HALL}^2 + \rho_{LONG}^2)$, where $\rho_{HALL} = R_{HALL}$, $\rho_{LONG} = R_{LONG} \times w/l$, $w$ is a Hall bar width, and $l$ is a distance between voltage probing contacts. **d**, Detailed investigation of the plateau for the filling factor $v$ = -2. The upper plot shows vanishing of the longitudinal resistance, and the lower plot shows precise coincidence of the measured quantum Hall resistance with the theoretical value of $-h/2e^2$, where $h$ and $e$ are Planck's constant and electronic charge value, respectively. The relative Hall resistance difference $\delta_r$ is defined as $\delta_r = (R_{HALL} - R_K/v) / R_K/v$ and yields only $(-3.5 \pm 3.7) \times 10^{-5}$.

**Figure 5.** Response of the heterostructure devices to water environment. **a**, The electric field-effect for a H-device at ambient conditions and with a water droplet on the heterostructure area, see inset in (**b**); arrows indicate the direction of voltage scans. At ambient conditions and in water a hysteresis in the gate effect between the subsequent opposite direction gate voltage scans was observed with a magnitude of 10V and 20V, respectively. This effect is most likely coursed by the trapped charges in devices at these conditions. **b**, Dynamic electrical response of the device upon placing and removing a water droplet on the heterostructures area (see main text for more details).

# Supporting Information (SI)

## All-carbon vertical van der Waals heterostructures:
## Non-destructive functionalization of graphene for electronic applications

Mirosław Woszczyna[1], Andreas Winter[2], Miriam Grothe[1], Annika Willunat[1], Stefan Wundrack[1], Rainer Stosch[1], Thomas Weimann[1], Franz Ahlers[1], and Andrey Turchanin[1,2*]

[1]*Physikalisch-Technische Bundesanstalt, 38116 Braunschweig, Germany*
[2]*Faculty of Physics, University of Bielefeld, 33615 Bielefeld, Germany*

e-mail: turchanin@physik.uni-bielefeld.de

## MATERIALS AND METHODS

### Production and Transfer of CVD graphene

The single-layer graphene (SLG) sheets were grown on 25 µm thick Cu-foils (Alfa Aesar, 99.8 %) cut into 2 x 2 cm² pieces in a tube furnace (Gero F40-200) by low-pressure chemical vapor deposition of methane (purity 4.5). The quartz glass tube of the furnace was loaded with copper foils, evacuated to $1 \times 10^{-3}$ mbar with a rotary pump (Edwards RV5) and filled with hydrogen (purity 5.3, ~1 mbar) under 50 sccm flow and heated to 1015°C with a rate of 150°C/h. The crystallinity of the copper foils had been improved by annealing for 30 min at 1015°C and then the hydrogen flow was reduced to 10 sccm and 70 sccm of methane were introduced at an overall pressure of ~2 mbar for 15 min. Then the furnace was cooled with a rate of ~250°C/min to ~250°C and the gas flow was stopped. Since graphene grows on both sides of the copper foil, the film from one side was protected with a layer of poly(methyl methacrylate) (PMMA, ~300 nm; All Resist, AR-P 679.04) and the graphene film on the opposite side was etched in an oxygen plasma for 60 s.

The graphene films were transferred from the Cu foils by etching in an aqueous solution (~3.5 %) of ammonium persulfate (Sigma Aldrich, 98%) overnight. Then the PMMA/graphene sandwich was fished out from the etching solution with a piece of silicon wafer and transferred on water to clean it from the etching solution for ~15 min and subsequently transferred onto the target substrate. Finally the PMMA was dissolved at room temperature in acetone (p.a.) overnight.

### Production and Transfer of NH$_2$-CNMs

NH$_2$-CNMs were prepared by electron-beam-induced crosslinking of 4´-nitro-1,1´-biphenyl-4-thiol (NBPT; Taros 99%) self-assembled monolayers (SAMs) on gold. To form the SAMs, we used 300 nm thermally evaporated Au on mica substrates (Georg

Albert PVD-Coatings). The substrates were cleaned in a UV/ozone-cleaner (FHR), rinsed with ethanol and blown dry in a stream of nitrogen. They were then immersed in a ~10 mmol solution of NBPT in dry, degassed dimethylformamide (DMF) for 72h in a sealed flask under nitrogen. Afterwards samples were rinsed with DMF and ethanol and blown dry with nitrogen. Cross-linking was achieved in high vacuum (<$5*10^{-7}$ mbar) with an electron floodgun (Specs) at an electron energy of 100 eV and a dose of 50 mC/cm$^2$.

The NH$_2$-CNMs were transferred onto a target substrate using a protecting layer of PMMA (AR-P 671.04) dissolved in chlorobenzene or ethyl acetate. This layer was used for mechanical stabilization of the nanosheets during the transfer process. Two layers of this polymer of overall thickness of ~400 nm were spun in sequence onto the CNM. First, a layer of low molecular weight PMMA (50 K) then a layer of high molecular weight PMMA (950 K) were spin-cast each for 30 s at 4000 rpm and cured on a hot plate at 90°C for 5 min. The underlying mica support was separated from the gold/cross-linked SAM/PMMA structure by a slight dipping into water of one of the edges/corners of the multilayered sample that finally (after separation) floats onto the air/water interface. Further, the sample was transferred by using a mica piece from the water surface to an I$_2$/KI/H$_2$O etching bath (1:4:10) where the gold film was dissolved within 15 min. Then the CNM was transferred to pure water for complete cleaning of the membrane from iodine contamination. Finally, the NH$_2$-CNM/PMMA structure was fished out by the target substrate (an oxidized silicon wafer with a large-area graphene sheet or the Hall bar structures) and the PMMA layer was dissolved in acetone.

### *X-ray Photoelectron Spectroscopy (XPS)*

A multi-chamber UHV-system (Multiprobe, Omicron) equipped XPS was employed for the analysis of the samples. X-ray photoelectron spectra were recorded using a monochromatic X-ray source (Al $K_\alpha$) and an electron analyzer (Sphera) with a resolution of 0.9 eV.

### *Raman Spectroscopy*

Raman spectra were acquired using a micro Raman spectrometer (LabRAM ARAMIS) operated in the backscattering mode. Measurements at 532 nm were obtained with a frequency-doubled Nd:YAG-Laser, a 50x long working distance objective (NA 0.50) and a thermoelectrically cooled CCD detector (2-3 $cm^{-1}$ spectral resolution). The spatial Raman imaging was performed with a motorized sample stage from Märzhäuser Wetzlar across an area of 40 x 145 µm and with a pixel size of 0.5 µm. During this measurement, the laser power and exposure time were kept very low to avoid local heating of the graphene sample.

### *Atomic Force Microscopy (AFM)*

AFM measurements were performed with Ntegra (NT-MDT) in contact mode at the ambient conditions using Pt-coated cantilevers (Phoenix Nanotechnologies) with the force constant of 0.03-0.20 N/m and the tip radius of 30-50 nm.

### *Fabrication of Hall Bar Structures (G- and H-Devices)*

Graphene Hall bars in graphene (G-devices) on Si/$SiO_2$-wafers (As-doped, resistivity 3-7 mΩcm, with 300 nm of the thermally grown silicon oxide, Si(100)) were fabricated using standard electron beam lithography and PMMA masks. Geometrical definition of the graphene shapes was achieved by dry etching in argon/oxygen plasma. Ti/Au contacts (10 nm/100 nm) were made by thermal evaporation and lift-off. To fabricate heterostructure devices (H-devices) the whole area of the 5 mm × 5 mm chips was covered with a $NH_2$-CNM, and the bonding Au wires were connected to the bonding through the encapsulation layer. As the pads were placed far away from the H-device area, the water droplets can easily be applied and removed without having contacts with non-insulated bonding Au wires. For the transport measurements the chips with devices were mounted onto 10 mm × 10 mm printed circuit boards using silver epoxy glue.

***Electric and Electromagnetic Transport Measurements***

Electrical transport measurements were carried out in an Oxford Instruments Helium 3 refrigerator HelioxTL. Before cooling down to the base temperature of 0.3 K each sample was kept in vacuum at $10^{-6}$ mbar for 1-24 hours at room temperature and subsequently electrically characterized. Keithley 2400 SourceMeter instruments were utilized to apply direct current and back-gate voltage. Voltages were measured by Keithley 2182A Nanovoltmeters. All field-effect devices were measured at a 1 µA direct source-drain current, $I_{DC}$. The mobility values were calculated as $\mu = (\rho n e)^{-1}$. Graphene resistivity, $\rho$, and charge carrier concentration, $n$, were expressed as $\rho = w/l \times V_{4T}/I_{DC}$ and $n = \alpha \times (V_{BG} - V_0)$, where $e$, $w$ and $l$ are elementary charge, device width and distance between two voltage probing contacts. The constant $\alpha = 7.2 \times 10^{10}$ cm$^{-2}$/V relates the induced graphene carrier concentration with the back-gate voltage $V_{BG}$ lowered by the charge neutrality voltage $V_0$, and applied to a 300 nm thick silicon oxide layer.

To estimate an effective interfacial potential induced by a water droplet at the water/$NH_2$-CNM interface after the protonation of amino groups (-$NH_3^+$) we applied the simple plate capacitor model. The protonated amino groups together with the 1 nm thick CNM dielectric create a top gate, which dopes graphene by electrons and shifts the charge neutrality point from the back-gate side by $\Delta V_{BG}$ = 27 Volts (see Fig. 5 of the article). The effective interfacial potential $\Delta V_{EF}$ can be then calculated using the formula $\Delta V_{EF} = \alpha \Delta V_{BG} t_{CNM} e / \varepsilon_0 \varepsilon_{CNM}$, where $\varepsilon_0$ is the vacuum permittivity, $t_{CNM}$ is the thickness of the CNM layer, $\varepsilon_{CNM}$ is its dielectric constant ($\varepsilon_{CNM}$ = 2.9). With these data we obtain $\Delta V_{EF} \approx 120$ mV.

**SI Table 1 │ Details of the Raman spectra presented in Figure 2b of the paper.** Raman spectra acquired from the bare SLG areas and the NH$_2$-CNM/SLG/SiO$_2$ areas of the sample presented in Figure 2a. The I(D)/I(G) ratio for both areas is similar showing that no additional structural defects are introduced into graphene while fabricating the heterostructure. The G-, 2D-peak positions, FWHM and intensity ratios of bare SLG and the NH$_2$-CNM/SLG/SiO$_2$ are averaged over 15.000 spectra (see SI Figure 1) and reflect a lower degree of doping impurities in the heterostructure in comparison to the bare graphene (see Refs. 19, 20 of the paper). Thus, the standard deviation of the FWHM of the G-peak for the heterostructure is ~50% lower in comparison to that for the bare graphene region.

| Material | ω(G), cm$^{-1}$ <br> FWHM(G), cm$^{-1}$ | ω(2D), cm$^{-1}$ <br> FWHM(2D), cm$^{-1}$ | I(D)/I(G) | I(2D)/I(G) |
|---|---|---|---|---|
| **SLG/SiO$_2$** | 1588.7 ± 1.2 <br> 14.4 ± 2.5 | 2680.3 ± 1,2 <br> 27.9 ± 1,8 | 0.12 ± 0.03 | 3.16 ± 0.62 |
| **NH$_2$-CNM/SLG/SiO$_2$** | 1588.8 ± 1.3 <br> 14.7 ± 1.3 | 2681.7 ± 1,2 <br> 30.2 ± 1,8 | 0.13 ± 0.03 | 2.87 ± 0.39 |

**SI Table 2 │ Details of the X-ray photoelectron spectroscopy (XPS) data presented in Figure 2c of the main paper.**

| Material | S2p$_{3/2}$/S2p$_{1/2}$, eV | C1s (FWHM), eV | N1s, eV |
|---|---|---|---|
| **SLG/SiO$_2$** | – | 284.7 (1.2) | – |
| **NH$_2$-CNM/SLG/SiO$_2$** | 164.5 / 165.7 | 284.7 (1.8) | 399.3 (-NH$_2$) <br> 401.2 (-NH$_3^+$) |

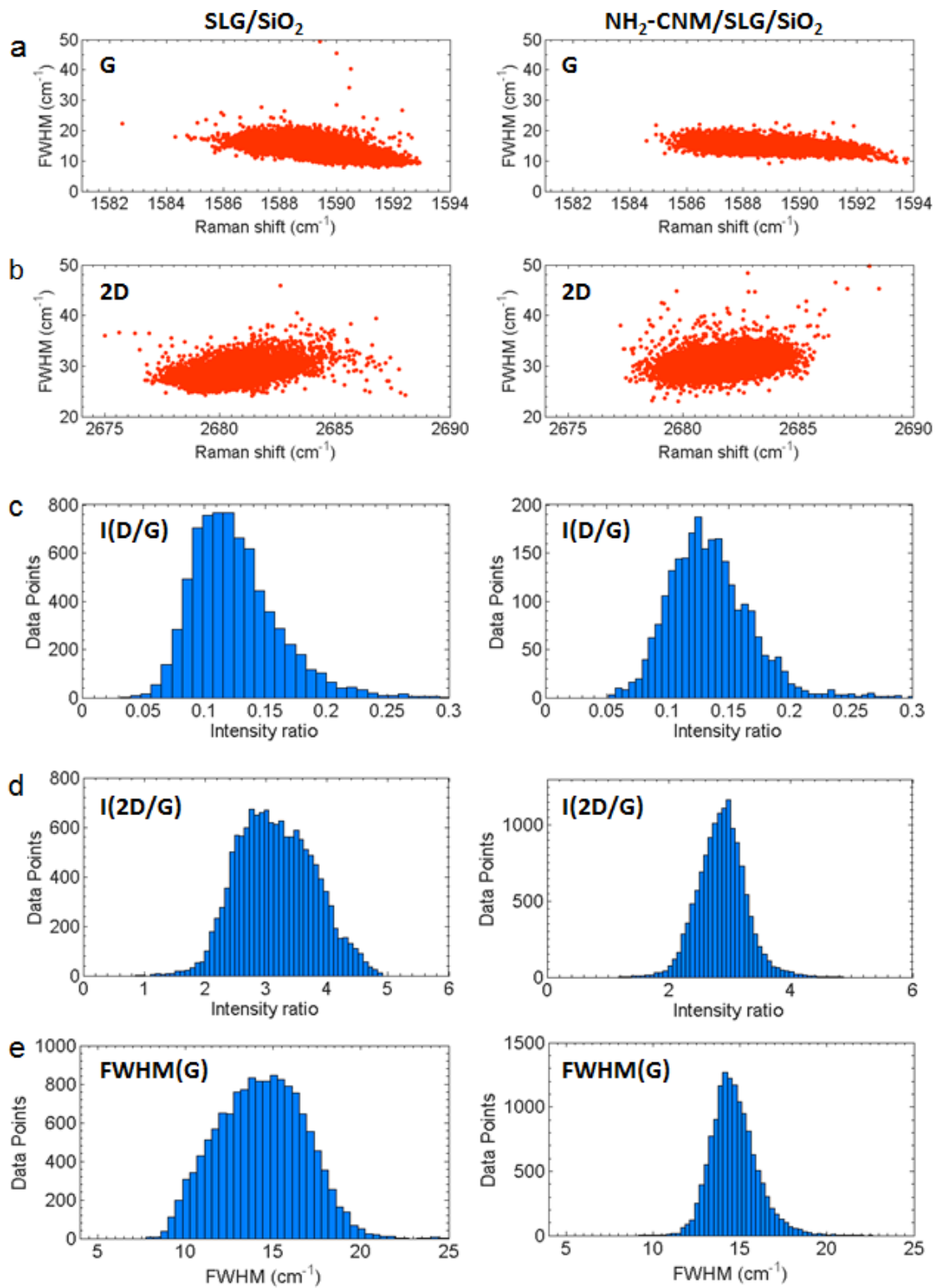

**SI Figure 1 | Detailed statistical analysis of Raman mappings from bare graphene and heterostructure a-b**, FWHM of G- and 2D-Peak as a function of Raman shift. **c,** I(D/G) distribution of $NH_2$-CNM/SLG/$SiO_2$ exhibits a similar low defect concentration in comparison to the bare graphene and represents its high structural quality. **d-e**, I(2D/G) and FWHM(G) distribution of bare graphene and heterostructure. An

increased number of small I(2D/G) ratios (~2.5) and low FWHM(G) (~11 cm$^{-1}$) indicates a higher degree of doping impurities in bare graphene. The narrow distributions of I(2D/G) ratios (~2.9) and FWHM(G) (~15 cm$^{-1}$) in the heterostructure demonstrate a lower degree of doping impurities and shows its encapsulating behavior against additional doping impurities.

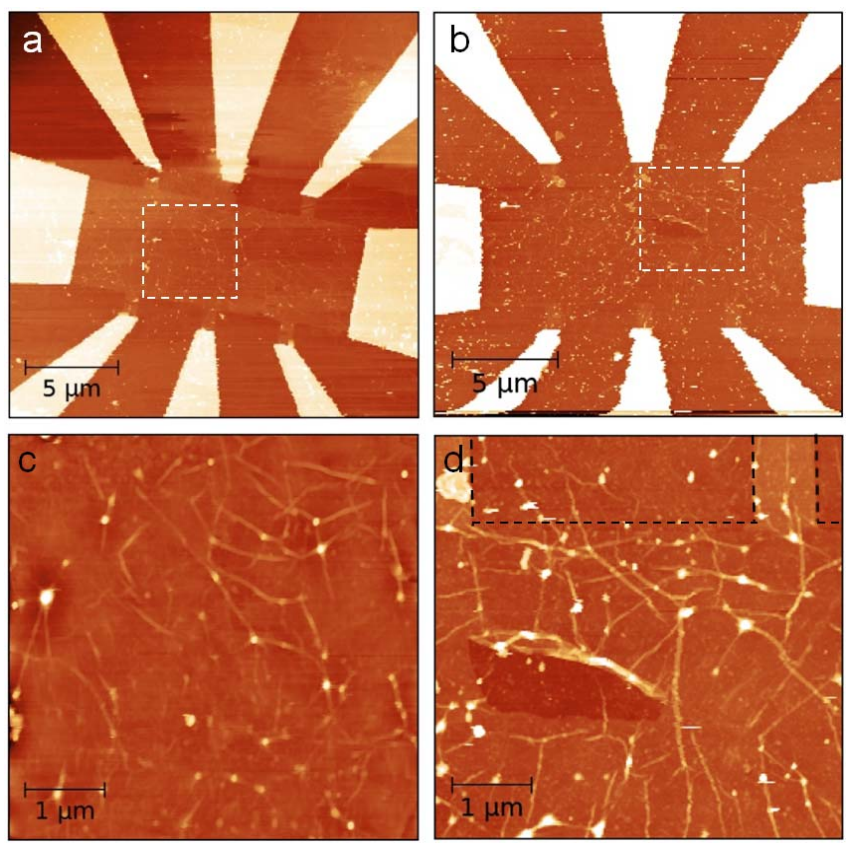

**SI Figure 2 | AFM images of G- and H-devices with dimensions as used for the "water droplet" experiment.** In (**a**) the low magnification image of a G-device with 100 nm thick gold contacts is presented (z-scale: 130 nm). Edges of the graphene area can clearly be recognized. Figure (**c**) shows the high magnification image of the area presented by the dashed square in (**a**) (z-scale: 50 nm). The folds can be seen, which are typical for the CVD graphene on silicon oxide substrate. The root-mean-square (RMS) roughness of the 3 µm × 3 µm area containing the folds is ~2.5 nm, whereas the RMS roughness from the smaller 1 µm × 1 µm nearly fold-free area is ~1.2 nm. The latter value corresponds well to the roughness of the bare silicon oxide surface. In (**b**) the low magnification image of an H-device is shown (z-scale: 90 nm). Due to complete coverage of this area with a $NH_2$-CNM, the graphene edges cannot be so well recognized as in (**a**). However, they are well seen in (**d**), where the high magnification image, corresponding to the dashed square in (**b**), is shown (z-scale: 40 nm). The black dashed lines in (**d**) are guiding the eye to the graphene edges. In the middle of (**d**) the $NH_2$-CNM is ruptured showing directly the underlying graphene surface. The density of folds in the heterostructure is increased in comparison to bare

graphene (compare (**c**) and (**d**)), as it results here from a superposition of the folds in both graphene and $NH_2$-CNM sheets. It can clearly be seen in (**d**) that the $NH_2$-CNM expands beyond the graphene area. The RMS roughness of the heterostructure for the 3 μm × 3 μm area is ~3.4 nm. In the nearly fold-free 1 μm × 1 μm area, the RMS roughness of the heterostructure is similar to the silicon oxide surface with a value of ~1.2 nm.